\documentclass{article}
\usepackage{spconf,amsmath,graphicx}
\usepackage{amssymb}
\usepackage{booktabs}
\usepackage{multirow}
\usepackage{hyperref}
\title{Deep Contextualized Acoustic Representations For\\ Semi-Supervised Speech Recognition}
\name{Shaoshi Ling, Yuzong Liu, Julian Salazar, Katrin Kirchhoff}
\address{Amazon AWS AI\\
\normalsize \texttt{\{shaosl,liuyuzon,julsal,katrinki\}@amazon.com}}
\begin{document}
\ninept
\maketitle
\begin{abstract}

We propose a novel approach to semi-supervised automatic speech recognition (ASR). We first exploit a large amount of unlabeled audio data via representation learning, where we reconstruct a temporal slice of filterbank features from past and future context frames. The resulting \textit{deep contextualized acoustic representations} (DeCoAR) are then used to train a CTC-based end-to-end ASR system using a smaller amount of labeled audio data. In our experiments, we show that  systems trained on DeCoAR consistently outperform ones trained on conventional filterbank features, giving 42\% and 19\% relative improvement over the baseline on WSJ \textit{eval92} and LibriSpeech \textit{test-clean}, respectively. Our approach can drastically reduce the amount of labeled data required; unsupervised training on LibriSpeech then supervision with 100 hours of labeled data achieves performance on par with training on all 960 hours directly. Pre-trained models and code will be released online.

\end{abstract}


\section{Introduction}
\label{sec:intro}

Current state-of-the-art models for speech recognition require vast amounts of transcribed audio data to attain good performance. In particular, end-to-end ASR models are more demanding in the amount of training data required when compared to traditional hybrid models. While obtaining a large amount of labeled data requires substantial effort and resources, it is much less costly to obtain abundant unlabeled data. 

For this reason, \textit{semi-supervised learning} (SSL) is often used when training ASR systems. The most commonly-used SSL approach in ASR is self-training \cite{thomas2013deep, huang2016semi, manohar2018semi, karita2018semi, parthasarathi2019lessons}. In this approach, a smaller labeled set is used to train an initial seed model, which is applied to a larger amount of unlabeled data to generate hypotheses. The unlabeled data with the most reliable hypotheses are added to the training data for re-training. This process is repeated iteratively. However, self-training is sensitive to the quality of the hypotheses and requires careful calibration of the confidence measures. Other SSL approaches include: pre-training on a large amount of unlabeled data with restricted Boltzmann machines (RBMs) \cite{vesely2013semi}; entropy minimization \cite{grandvalet2005semi, huang2010semi, yu2010active}, where the uncertainty of the unlabeled data is incorporated as part of the training objective; and graph-based approaches \cite{liu2016graph}, where the manifold smoothness assumption is exploited.  Recently, transfer learning from large-scale pre-trained language models (LMs) \cite{peters2018deep, devlin2019bert, yang2019xlnet} has shown great success and achieved state-of-the-art performance in many NLP tasks. The core idea of these approaches is to learn efficient word representations by pre-training on massive amounts of unlabeled text via word completion. These representations can then be used for downstream tasks with labeled data. 

Inspired by this, we propose an SSL framework that learns efficient, context-aware acoustic representations using a large amount of unlabeled data, and then applies these representations to ASR tasks using a limited amount of labeled data. In our implementation, we perform acoustic representation learning using forward and backward LSTMs and a training objective that minimizes the reconstruction error of a temporal slice of filterbank features given previous and future context frames. After pre-training, we fix these parameters and add output layers with connectionist temporal classification (CTC) loss for the ASR task.

The paper is organized as follows: in Section \ref{sec:related_work}, we give a brief overview of related work in acoustic representation learning and SSL. In Section \ref{sec:approach}, we describe an implementation of our SSL framework with DeCoAR learning. We describe the experimental setup in Section \ref{sec:expt} and the results on WSJ and LibriSpeech in Section \ref{sec:results}, followed by our conclusions in Section \ref{sec:conclusion}.

\begin{figure*}[t]
  \centering
  \includegraphics[width=0.9\linewidth]{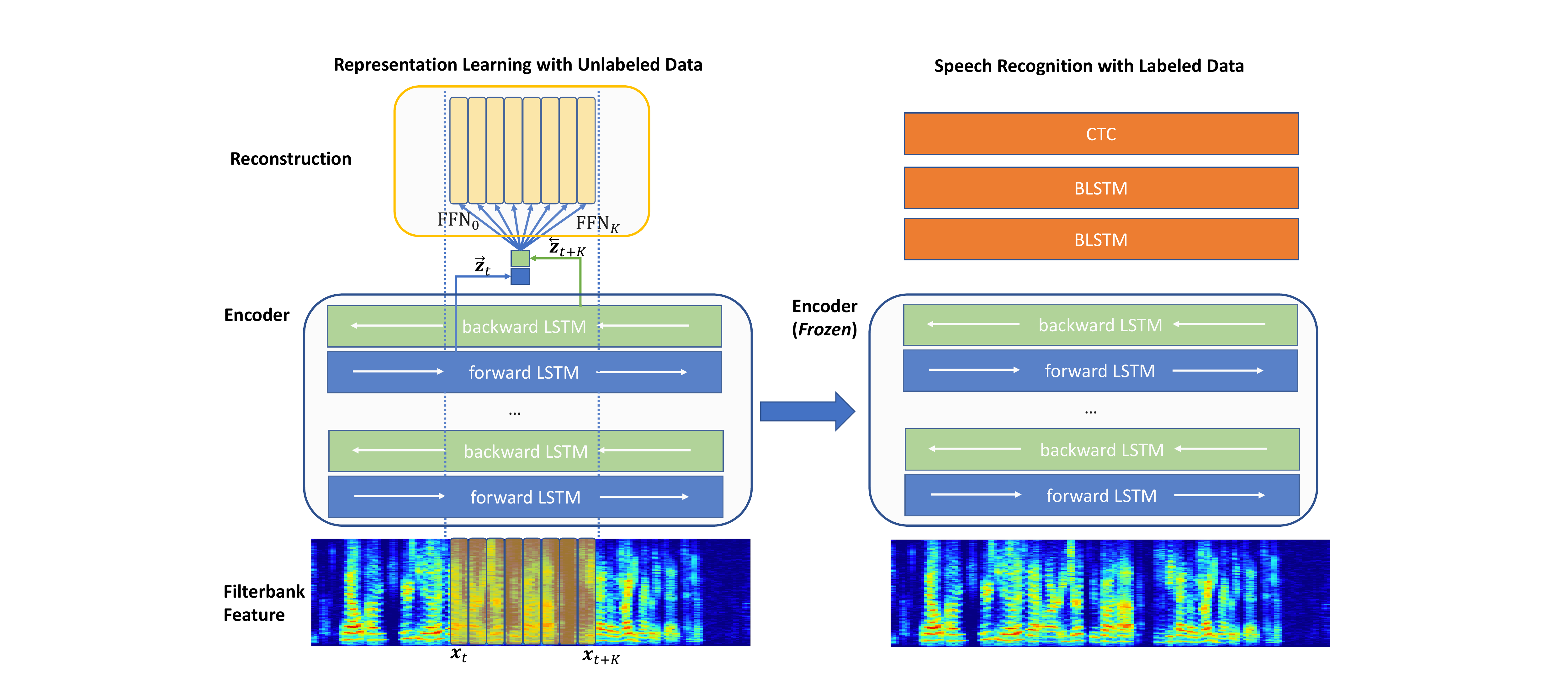}
  \caption{Illustration of our semi-supervised speech recognition system.}
  \label{fig:model_view}
\end{figure*}

\section{Related work}
\label{sec:related_work}

While semi-supervised learning has been exploited in a plethora of works in hybrid ASR system, there are very few work done in the end-to-end counterparts \cite{karita2018semi, Baskar2019, Dey2019}. In \cite{karita2018semi}, an intermediate representation of speech and text is learned via a shared encoder network. To train these representation, the encoder network was trained to optimize a combination of ASR loss, text-to-text autoencoder loss and inter-domain loss. The latter two loss functions did not require paired speech and text data.  Learning efficient acoustic representation can be traced back to restricted Boltzmann machine \cite{hinton2006fast, hinton2012deep, bengio2007greedy}, which allows pre-training on large amounts of unlabeled data before training the deep neural network acoustic models. 

More recently, acoustic representation learning has drawn increasing attention  \cite{hsu2018extracting, chung2018speech2vec, chung2019unsupervised, chorowski2019unsupervised, schneider2019wav2vec, baevski2019vq} in speech processing. For example, an autoregressive predictive coding model (APC) was proposed in \cite{chung2019unsupervised} for unsupervised speech representation learning and was applied to phone classification and speaker verification. WaveNet auto-encoders \cite{chorowski2019unsupervised} proposed contrastive predictive coding (CPC) to learn speech representations and was applied on unsupervised acoustic unit discovery task. Wav2vec~\cite{schneider2019wav2vec} proposed a multi-layer convolutional neural network optimized via a noise contrastive binary classification and was applied to WSJ ASR tasks. 

Unlike the speech representations described in \cite{schneider2019wav2vec, chung2019unsupervised}, our representations are optimized to use bi-directional contexts to auto-regressively reconstruct unseen frames. Thus, they are deep contextualized representations that are functions of the entire input sentence. More importantly, our work is a general semi-supervised training framework that can be applied to different systems and requires no architecture change. 

\section{DEep COntextualized Acoustic Representations}
\label{sec:approach}

\subsection{Representation learning from unlabeled data}

Our approach is largely inspired by ELMo \cite{peters2018deep}. In ELMo, given a sequence of $T$ tokens $(w_1,w_2,...,w_T)$, a forward language model (implemented with an LSTM) computes its probability using the chain rule decomposition:
\begin{equation*}
  \begin{aligned}
p(w_1,w_2,\cdots,w_T) = \prod_{t=1}^T p(w_k \mid w_1,w_2,\dotsc,w_{t-1}).
\end{aligned}
\end{equation*}
Similarly, a backward language model computes the sequence probability by modeling the probability of token $w_t$ given its future context $w_{t+1},\dotsc, w_T$ as follows:
\begin{equation*}
  \begin{aligned}
p(w_1,w_2,\cdots,w_T) = \prod_{t=1}^T p(w_t \mid w_{t+1},w_{t+2},\dotsc,w_T)
\end{aligned}
\end{equation*}

ELMo is trained by maximizing the joint log-likelihood of both forward and backward language model probabilities:
\begin{multline}
\sum_{t=1}^{T} [\log p(w_t \mid w_1,w_2,...,w_{t-1};\ \Theta_x,\overrightarrow{\Theta}_{\text{LSTM}},\Theta_s)\\+\log p(w_t \mid w_{t+1},w_{t+2},...,w_{T};\ \Theta_x,\overleftarrow{\Theta}_{\text{LSTM}},\Theta_s)]
\end{multline}
where $\Theta_x$ is the parameter for the token representation layer, $\Theta_s$ is the parameter for the softmax layer, and $\overrightarrow{\Theta}_{\text{LSTM}}$, $\overleftarrow{\Theta}_{\text{LSTM}}$ are the parameters of forward and backward LSTM layers, respectively. As the word representations are learned with neural networks that use past and future information, they are referred to as deep contextualized word representations.

For speech processing, the input audio data is represented as a sequence of acoustic features $\mathbf{X} = (\mathbf{x}_1,\cdots,\mathbf{x}_T)$ (e.g. filterbank or MFCC features). Predicting an acoustic feature $\mathbf{x}_t$ may be a trivial task, as it could be solved by exploiting the temporal smoothness of the signal. In the APC model \cite{chung2019unsupervised}, the authors propose to predict a frame $K$ steps ahead of the current one. Namely, the model aims to minimize the $\ell_1$ loss between an acoustic feature vector $\mathbf{x}$ at time $t+K$ and a reconstruction $\mathbf{y}$ predicted at time $t$: $\sum_{t=1}^{T-K} |\mathbf{x}_{t+K} - \mathbf{y}_t|$. They conjectured this would induce the model to learn more global structure rather than simply leveraging local information within the signal.

We propose combining the bidirectionality of ELMo and the reconstruction objective of APC to give \textit{deep contextualized acoustic representations} (DeCoAR). We train the model to predict a \textit{slice} of acoustic feature vectors, given past and future acoustic vectors. As depicted on the left side of Figure~\ref{fig:model_view}, a stack of forward and backward LSTMs are applied to the entire unlabeled input sequence $\mathbf{X} = (\mathbf{x}_1,\cdots,\mathbf{x}_T)$, where the $\mathbf{x}_t$ are 40-dimensional log filterbank features. The network computes a hidden representation that encodes information from both previous and future frames (i.e. $\overrightarrow{\mathbf{z}}_t, \overleftarrow{\mathbf{z}}_t$) for each frame $\mathbf{x}_t$.  Given a sequence of acoustic feature inputs  $(\mathbf{x}_1, ..., \mathbf{x}_{T}) \in \mathbb{R}^d$, for each slice $(\mathbf{x}_t, \mathbf{x}_{t+1}, ..., \mathbf{x}_{t+K})$ starting at time step $t$, our objective is defined as follows:
\begin{equation}\label{eq:decoar}
  \begin{aligned}
\mathcal{L}_t= \sum_{i=0}^{K}|\mathbf{x}_{t+i} - \text{FFN}_i ([\overrightarrow{\mathbf{z}}_t; \overleftarrow{\mathbf{z}}_{t+K}])|
\end{aligned}
\end{equation}
where $[\overrightarrow{\mathbf{z}}_t; \overleftarrow{\mathbf{z}}_{t}] \in \mathbb{R}^{2h}$ are the concatenated forward and backward states from the last LSTM layer, and
\begin{equation}
    \text{FFN}_i(\mathbf{v}) = \mathbf{W}_{i,2} \text{ReLU}(\mathbf{W}_{i,1} \mathbf{v} + \mathbf{b}_{i,1})  + \mathbf{b}_{i,2}
\end{equation}
is a position-dependent feed-forward network with 512 hidden dimensions. The final loss $\mathcal{L}$ is summed over all possible slices in the entire sequence:
\begin{equation}
  \begin{aligned}
\mathcal{L}=\sum_{t=1}^{T-K}\mathcal{L}_t
\end{aligned}
\end{equation}
Note this can be implemented efficiently as a layer which predicts these $(K+1)$ frames at each position $t$, all at once. We compare with the use of unidirectional LSTMs and various slice sizes in Section \ref{sec:results}.

\begin{table*}[th]
    \centering
   \begin{tabular}{ c |cc|cc|cc|cc }
       \toprule
        \multirow{2}{*}{\textbf{Representation}} & \multicolumn{2}{c|}{\textbf{100 hours}} & \multicolumn{2}{c|}{\textbf{360 hours}} & \multicolumn{2}{c|}{\textbf{460 hours}} & \multicolumn{2}{c}{\textbf{960 hours}} \\
         & test-clean & test-other & test-clean & test-other & test-clean & test-other & test-clean & test-other \\ 
        \midrule
        filterbank                          & 9.36 & 30.20 &   7.57 & 25.28 & 7.11 & 24.31  & 5.82 & 14.50 \\
        wav2vec \cite{schneider2019wav2vec} & 6.92 &  20.00 & 6.26 & 18.17  & 6.01 & 17.00   & 5.12 &  13.07\\
        DeCoAR                              & 6.10 &17.43 & 5.23 & 14.67   & 5.12 & 14.10  & 4.74 & 12.20 \\
        \bottomrule
    \end{tabular}
    \caption{Semi-supervised LibriSpeech results.}
    \label{table:LibriSpeech}
\end{table*}

\subsection{End-to-end ASR training with labeled data}

After we have pre-trained the DeCoAR on unlabeled data, we freeze the parameters in the architecture. To train an end-to-end ASR system using labeled data, we remove the reconstruction layer and add two BLSTM layers with CTC loss \cite{graves2016ctc}, as illustrated on the right side of Figure~\ref{fig:model_view}. The DeCoAR vectors induced by the labeled data in the forward and backward layers are concatenated. We fine-tune the parameters of this ASR-specific new layer on the labeled data. 

While we use LSTMs and CTC loss in our implementation, our SSL approach should work for other layer choices (e.g. TDNN, CNN, self-attention) and other downstream ASR models (e.g. hybrid, seq2seq, RNN transducers) as well.

\section{Experimental Setup}
\label{sec:expt}

\subsection{Data}
We conducted our experiments on the WSJ and LibriSpeech datasets, pre-training by using one of the two training sets as unlabeled data. To simulate the SSL setting in WSJ, we used 30\%, 50\% as well as 100\% of labeled data for ASR training, consisting of 25 hours, 40 hours, and 81 hours, respectively. We used \textit{dev93} for validation and \textit{eval92} and evaluation. For LibriSpeech, the amount of training data used varied from 100 hours to the entire 960 hours. We used \textit{dev-clean} for validation and \textit{test-clean}, \textit{test-other} for evaluation.

\subsection{ASR systems}
Our experiments consisted of three different setups: 1) a fully-supervised system using all labeled data; 2) an SSL system using wav2vec features; 3) an SSL system using our proposed DeCoAR features.  All models used were based on deep BLSTMs with the CTC loss criterion. 

In the supervised ASR setup, we used conventional log-mel filterbank features, which were extracted with a 25ms sliding window at a 10ms frame rate. The features were normalized via mean subtraction and variance normalization on a per-speaker basis.  The model had 6 BLSTM layers, with 512 cells in each direction. We found that increasing the number of cells to a larger number did not further improve the performance and thus used it as our best supervised ASR baseline.  The output CTC labels were 71 phonemes\footnote{The CMU lexicon: \url{http://www.speech.cs.cmu.edu/cgi-bin/cmudict}} plus one blank symbol.

In the SSL ASR setup, we pre-trained a 4-layer BLSTM (1024 cells per sub-layer) to learn DeCoAR features according to the loss defined in Equation \ref{eq:decoar} and use a slice size of 18. We optimized the network with SGD and use a Noam learning rate schedule, where we started with a learning rate of 0.001, gradually warm up for 500 updates, and then perform inverse square-root decay. We grouped the input sequences by length with a batch size of 64, and trained the models on 8 GPUs. After the representation network was trained, we froze the parameters, and added a projection layer, followed by 2-layer BLSTM with CTC loss on top it. We fed the labeled data to the network. For comparison, we obtained 512-dimensional wav2vec representations \cite{schneider2019wav2vec} from the wav2vec-large model\footnote{\url{https://github.com/pytorch/fairseq/tree/master/examples/wav2vec}}. Their model was pre-trained on 960-hour LibriSpeech data with constrastive loss and had 12 convolutional layers with skip connections. 

For evaluation purposes, we applied WFST-based decoding using EESEN \cite{miao2015eesen}. We composed the CTC labels, lexicons and language models (unpruned trigram LM for WSJ, 4-gram for LibriSpeech\footnote{Downloaded from \url{http://www.openslr.org/11}}) into a decoding graph. The acoustic model score was set to $0.8$ and $1.0$ for WSJ and LibriSpeech, respectively, and the blank symbol prior scale was set to $0.3$ for both tasks. We report the performance in word error rate (WER).

\begin{table}[th]
    \centering
    \ninept
    \begin{tabular}{  ccc|cc  }
        \toprule 
        \textbf{Representation} & \textbf{Unlabeled} & \textbf{Labeled} & \textbf{dev93} & \textbf{eval92} \\
        \midrule
        filterbank &-& 81h & 8.21 & 5.44\\
        wav2vec \cite{schneider2019wav2vec} & 960h Libri. & 81h & 6.84 & 3.97 \\
        DeCoAR & 960h Libri. & 81h & 6.30 & 3.17 \\
        \midrule
        filterbank & - & 25h & 18.16 & 11.04 \\
        filterbank & - & 40h & 13.50 & 9.20 \\
        DeCoAR & 81h WSJ & 25h & 10.38 & 5.81 \\
        DeCoAR & 81h WSJ & 40h & 9.41 & 5.09 \\
        DeCoAR & 81h WSJ & 81h & 8.34 & 4.64 \\
        \bottomrule
    \end{tabular}
    \caption{Semi-supervised WSJ results. \textit{Unlabeled} indicates the amount of unlabeled data used for acoustic representation learning, and \textit{Labeled} indicates the amount of labeled data in ASR training.}
    \label{table:wsj}
\end{table}

\begin{figure*}[t]
  \centering
  \includegraphics[width=0.9\linewidth]{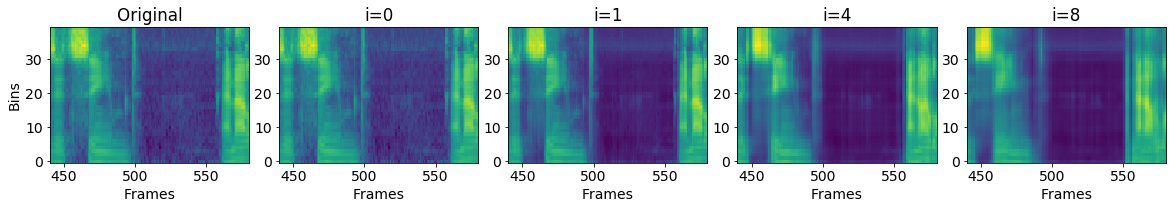}
  \caption{The spectrograms for a portion of LibriSpeech \textit{dev-clean} utterance 2428-83699-0034 reconstructed by generated by taking the $i$-th frame prediction (slice size 18) at each time step. The reconstruction becomes less noisy but more simplistic when predicting further into the masked slice.}
  \label{fig:spectrogram}
\end{figure*}

\section{Results}
\label{sec:results}
\subsection{Semi-supervised WSJ results}
Table \ref{table:wsj} shows our results on semi-supervised WSJ. We demonstrate that DeCoAR feature outperforms filterbank and wav2vec features, with a relative improvement of 42\% and 20\%, respectively. The lower part of the table shows that with smaller amounts of labeled data, the DeCoAR features are significantly better than the filterbank features: Compared to the system trained on 100\% labeled data with filterbank features, we achieve comparable results on \textit{eval92} using 30\% of the labeled data and better performance on \textit{eval92} using 50\% of the labeled data. 

\subsection{Semi-supervised LibriSpeech results}
Table \ref{table:LibriSpeech} shows the results on semi-supervised LibriSpeech. Both our representations and wav2vec \cite{schneider2019wav2vec} are trained on 960h LibriSpeech data. We conduct our semi-supervised experiments using 100h (\textit{train-clean-100}), 360h (\textit{train-clean-360}), 460h, and 960h of training data. Our approach outperforms both the baseline and wav2vec model in each SSL scenario. One notable observation is that using only 100 hours of transcribed data achieves very similar performance to the system trained on the full 960-hour data with filterbank features. On the more challenging \textit{test-other} dataset, we also achieve performance on par with the filterbank baseline using a 360h subset. Furthermore, training with with our DeCoAR features approach improves the baseline even when using the exact same training data (960h). Note that while \cite{park2019specaugment} introduced SpecAugment to significantly improve LibriSpeech performance via data augmentation, and \cite{luscher2019rwth} achieved state-of-the-art results using both hybrid and end-to-end models, our approach focuses on the SSL case with less labeled training data via our DeCoAR features.

\subsection{Ablation Study and Analysis}

\subsubsection{Context window size}
We study the effect of the context window size during pre-training. Table~\ref{table:window_size} shows that masking and predicting a larger slice of frames can actually degrade performance, while increasing training time. A similar effect was found in SpanBERT \cite{joshi2019spanbert}, another deep contextual word representation which found that masking a mean span of 3.8 consecutive words was ideal for their word reconstruction objective.
\begin{table}[th]
    \centering
    \ninept
    \begin{tabular}{ c | cc  }
        \toprule
        {\textbf{Slice size}} & {\textbf{dev93} } & {\textbf{eval92} } \\
        \midrule
        12 &  6.58 &  3.50\\
        18 &  6.30 & 3.17\\
        22 &  6.62 & 3.62\\
        \bottomrule
    \end{tabular}
    \caption{Comparison of WERs on WSJ after pre-training with different slice window sizes ($K+1$) on LibriSpeech.}
    \label{table:window_size}
\end{table}

\subsection{Unidirectional versus bidirectional context}

Next, we study the importance of bidirectional context by training a unidirectional LSTM, which corresponds to only using $\overrightarrow{\mathbf{z}}_t$ to predict $\mathbf{x}_t, \dotsc, \mathbf{x}_{t+K}$. Table~\ref{table:bidirection} shows that this unidirectional model achieves comparable performance to the wav2vec model \cite{schneider2019wav2vec}, suggesting that bidirectionality is the largest contributor to DeCoAR's improved performance.
\begin{table}[th]
    \centering

    \ninept
    \begin{tabular}{ cc|cc  }
        \toprule
        \textbf{Representation} & \textbf{Context} & \textbf{dev93} & \textbf{eval92} \\
        \midrule
        filterbank & -- & 8.21 & 5.44 \\
        wav2vec & unidirectional &  6.84 & 3.97   \\
        DeCoAR & unidirectional & 6.87 & 3.62  \\
        DeCoAR &  bidirectional & 6.30  & 3.17\\
        \bottomrule
    \end{tabular}
    \caption{Comparison of WERs on WSJ after pre-training using unidirectional or bidirectional context on LibriSpeech.}
    \label{table:bidirection}
\end{table}

\subsubsection{DeCoAR as denoiser}
Since our model is trained by predicting masked frames, DeCoAR has the side effect of learning decoder feed-forward networks $\text{FFN}_i$ which reconstruct the $(t+i)$-th filterbank frame from contexts $\overrightarrow{\mathbf{z}}_t$ and $\overleftarrow{\mathbf{z}}_{t+K}$. In this section, we consider the spectrogram reconstructed by taking the output of $\text{FFN}_i$ at all times $t$.

The qualitative result is depicted in Figure~\ref{fig:spectrogram} where the slice size is 18. We see that when $i=0$ (i.e., when reconstructing the $t$-th frame from $[\overrightarrow{\mathbf{z}}_t; \overleftarrow{\mathbf{z}}_{t+K}]$), the reconstruction is almost perfect. However, as soon as one predicts unseen frames $i=1, 4, 8$ (of 16), the reconstruction becomes more  simplistic, but not by much. Background energy in the silent frames 510-550 is zeroed out. By $i=8$ artifacts begin to occur, such as an erroneous sharp band of energy being predicted around frame 555. This behavior is compatible with recent NLP works that interpret contextual word representations as denoising autoencoders \cite{yang2019xlnet}.

The surprising ability of DeCoAR to broadly reconstruct a frame $\overrightarrow{\mathbf{x}}_{t+{K/2}}$ in the middle of a missing 16-frame slice suggests that its representations $[\overrightarrow{\mathbf{z}}_t; \overleftarrow{\mathbf{z}}_{t+K}]$ capture longer-term phonetic structure during unsupervised pre-training, as with APC \cite{chung2019unsupervised}. This motivates its success in the semi-supervised ASR task with only two additional layers, as it suggests DeCoAR learns phonetic representations similar to those likely learned by the first 4 layers of a corresponding end-to-end ASR model.

\section{Conclusion}
\label{sec:conclusion}

In this paper, we introduce a novel semi-supervised learning approach for automatic speech recognition. We first propose a novel objective for a deep bidirectional LSTM network, where large amounts of unlabeled data are used to learn deep contextualized acoustic representations (DeCoAR). These DeCoAR features are used as the representations of labeled data to train a CTC-based end-to-end ASR model. In our experiments, we show a 42\% relative improvement on WSJ compared to a baseline trained on log-mel filterbank features. On LibriSpeech, we achieve similar performance to training on 960 hours of labeled by pretraining then using only 100 hours of labeled data. While we use BLSTM-CTC as our ASR model, our approach can be applied to other end-to-end ASR models.


\bibliographystyle{IEEEbib}
\bibliography{refs}

\end{document}